# Rigid3D: a hybrid multi-sphere DEM framework for simulation of non-spherical particles in multi-phase flow

Fei-Liang Yuan[1,2], Martin Sommerfeld[1], Pradeep Muramulla[2], Srikanth Gopireddy[2], Lars Pasternak[1], Nora Urbanetz[2], Thomas Profitlich[2]

[1] *Multiphase Flow Systems, Institute of Process Engineering, Otto-von-Guericke-University Magdeburg, Hoher Weg 7b, D-06120, Halle (Saale), Germany*
[2] *Daiichi-Sankyo Europe GmbH, Pharmaceutical Development, Luitpoldstrasse 1, 85276, Pfaffenhofen, Germany*

**Abstract**

This article presents the development and validation of a hybrid multi-sphere discrete element framework - Rigid3D, for the simulation of granular systems with arbitrarily shaped particles in 3D space. In this DEM framework, a non-spherical particle is approximated by three different geometric models: (1) multi-sphere model with overlapping spheres (MS model), (2) particle surface with triangle mesh (surface model), and (3) discretized particle body with polyhedral cells (cell model). The multi-sphere approach will be the "engine" for efficient DEM simulations, while particle's mesh and cell models will be updated simultaneously according to the position and orientation of their associated MS model, for use in particle related inter-phase couplings in a multi-phase flow. In this sense, Rigid3D tries to combine the best of both worlds in multi-sphere and polyhedral DEMs: multi-sphere method for the efficiency and acceptable accuracy in the DEM simulation of granular flows, while the surface and cell models for the couplings between particles and other phases (continuous or dispersed phases) without affecting the performance of DEM simulations.

## 1. Motivation

Multi-phase flows involving non-spherical particles are wildly encountered in many industrial processes. Due to the complexity of inter-phase interactions and limited accessibility of real facilities, it can be difficult to obtain insight information with measurements. Therefore, accurate numerical models that can provide micro- and macro-scale information, are desired tools for design and optimization purposes. Numerical studies of particle relevant inter-phase couplings usually require details of particle's surface and volumetric information, e.g. particle-fluid interaction, spray coating of pharmaceutical tablets, heat and mass transfer between solid particles and surrounding fluids, etc. Therefore, an efficient DEM framework with accurate representation of particle's geometry information will be necessary for these tasks.

Generally speaking, there are two versatile approaches to model particles with realistic shapes either convex or concave: multi-sphere (MS) DEM and polyhedral DEM, because super-quadrics, poly-superquadrics [4, 13, 25, 31, 40, 46] or any other mathematically described shapes usually have symmetrical, continuous and smooth surfaces, thus are not sufficient to represent realistic particles that are usually asymmetrical and angular. Polyhedral DEMs are probably the most straightforward approach to model particle's surface using triangle mesh, that can be directly obtained from 3D-scanner or meshing software. Nevertheless, this convenience in shape description does come at a price. First, expensive contact detection between particle surface elements: vertices, edges and faces (triangles) with six different types of contact pair, namely vertex-to-vertex, vertex-to-edge, vertex-to-face, edge-to-edge, edge-to-face and face-to-face. For non-convex particles, complex algorithms [e.g. 10, 17, 20, 28, 33, 45] usually need to be implemented instead of using conventional approaches [3, 5, 29] for convex shapes. Second, a few thousand of triangles are normally required for a good approximation of particles with arbitrary shapes, thus limits the number of particles to be simulated using polyhedral DEMs. For a coarse surface resolution of one thousand triangles per particle, 100 thousand particles would result in 100 million faces, which is computationally expensive as most users usually have access to only a few dozes of CPU cores. MS-DEM [9] uses spheres to approximate particle shapes [see e.g. 11, 14, 22, 44], thus the contact detection between particles can be simply handled with the sphere-sphere contact problem rather than the expensive triangle-triangle contact in polyhedral DEMs. As the particle shape in MS-DEM is literally the boolean union of all primary spheres, the surface of MS particle is usually bumpy and the curvature of surface is discontinuous. Therefore the collision on the level of a single pair of contacting particles in MS-DEM may not be very accurate compared to polyhedral DEMs. Nevertheless, for granular systems with many non-spherical particles, this side-effect may compensate each other, and the bulk properties such as packing porosity, static or dynamic angle of repose, and flow patterns, etc., from MS-DEM simulations can still converge to experimental results [see e.g. 16, 23].

Based on the investigation above, this article aims to present the implementation and validation of a hybrid DEM code - Rigid3D, where some of the best features from multi-sphere and polyhedral DEMs are incorporated into one framework. (1) multi-sphere (MS) model for efficient contact detection; (2) accurate polyhedral model for precise description of particle surface; (3) if particle's volumetric information is also needed, e.g., for particle-fluid coupling, then the cell model can be created from discretized particle body with small polyhedral cells. As particle's MS model is used for the DEM simulation of particle's motion, thus it is relatively simple for the code implementation, and the contact between particles can be simply handled with sphere-sphere contact problem.

For inter-phase coupling between solid particles and fluid phases where the information about particle's geometry in moderate to high resolution is required, this hybrid DEM framework is particularly efficient, as the particle's surface and cell models are not involved in the contact detection. This feature is crucial for the accuracy and efficiency of the particle related inter-phase couplings, as it is too expensive in polyhedral DEMs to approximate particles with surface mesh of high resolution (e.g. 5~10 thousand triangles) in granular systems with large number of non-spherical particles.

## 2. Hybrid DEM framework

This section will demonstrate the concept of the hybrid DEM framework Rigid3D using the multi-sphere approach as the "engine" of DEM simulation, while particle's surface and cell models are associated with their corresponding MS model for the inter-phase couplings. An ellipsoidal particle is used here for the sake of simplicity, in order to show cases of multi-sphere model construction, contact detection and neighbor list building. Since the essential ingredients for typical soft-sphere based DEM are already covered in SR-DEM (Surface of Revolution DEM framework) recently developed by the author, therefore theoretical background of DEM will be briefly outlined here for the sake of completeness.

### 2.1 Particle models

As mentioned earlier in the abstract, a non-spherical particle can have three geometric models in the Rigid3D code: multi-sphere (MS) model, surface model and cell model. The first one is mandatory as it is used in the contact detection, while the latter two are optional whenever they are needed. A particle constructed by "clumping" multiple spheres together will be termed as MS particle (i.e. MS model) in this work. The primary (or component) spheres in a MS particle are fixed in the body frame, i.e. their relative positions do not change during contacts. Multi-sphere approach provides great versatility in particle shape representation especially for smooth particles, and simplicity in code implementation.

MS particles are usually generated by filling maximum inscribed spheres (MIS) that have at least two closest points on the particle's boundary. An ellipse ($\frac{x^2}{a^2} + \frac{y^2}{b^2} = 1$, a = 2b) in Figure 1(a) can be roughly approximated by only 7 maximum inscribed circles shown in Figure 1(b). Since the MS model is literally the boolean union of all primary circles in 2D or primary spheres in 3D, the particle's boundary or surface is usually bumpy as can be seen in Figure 1(c-d), which implies the surface of a MS particle is always non-convex. With more primary spheres filling up the concave gaps on the surface, MS particle's surface can become smooth. As shown in Figure 2(b), an ellipsoidal particle ($\frac{x^2}{a^2} + \frac{y^2}{b^2} + \frac{z^2}{c^2} = 1$, $a = 2b = 2c$) can be well approximated by 15 primary spheres, while the coarse model with 7 primary spheres has rather bumpy surface in Figure 2(a).

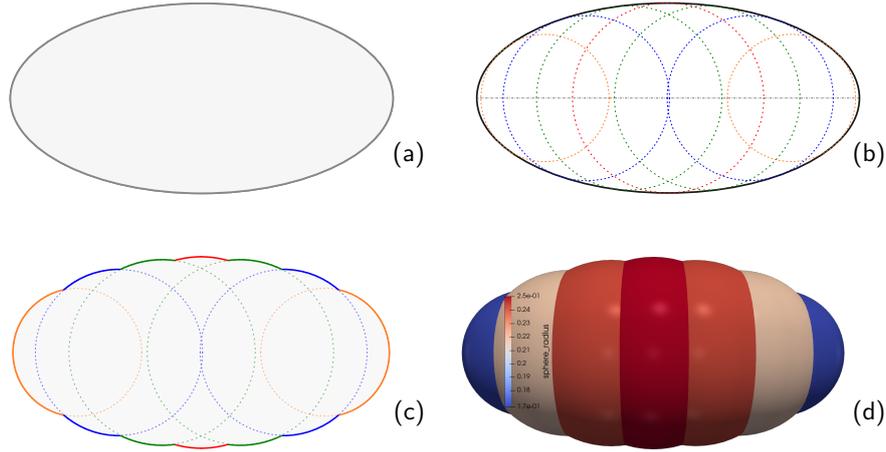

**Figure 1:** Schematic of 2D/3D multi-sphere models. (a) original particle - ellipse; (b) approximation with 7 maximum inscribed circles; (c) the real surface of 2D MS particle with connected arcs from inscribed circles; (d) corresponding 3D MS particle with 7 spheres.

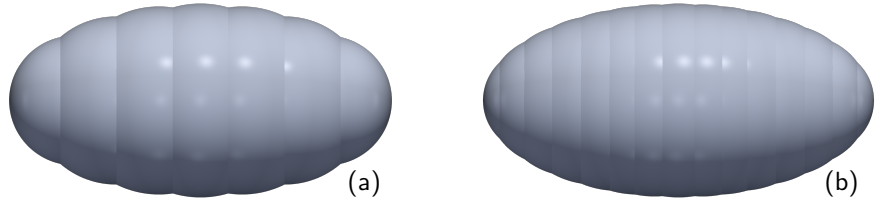

**Figure 2:** MS models of ellipsoidal particle with intermediate and fine resolution. (a) bumpy surface with 7 primary spheres; (b) relatively smooth surface with 15 primary spheres.

One drawback associated with particle's MS model is the difficulty to access the details of surface information, even if the original particle shape can be accurately assembled by primary spheres. Because a sphere is implicitly defined by its center and radius, therefore the surface of MS particle, i.e. boolean union of all primary spheres, is also implicit. In case of inter-phase couplings, e.g. spray droplets deposition on a pharmaceutical tablet surface and particle-fluid interaction, the details about the particle surface and volumetric information are usually required. Therefore two extra geometric models, i.e. the surface and cell models are linked with the particle's MS model as shown in Figure 3, and their position and orientation are also updated accordingly with the corresponding MS model at each coupling interval.

The hybrid DEM framework Rigid3D using multiple geometric models has several advantages over traditional polyhedral DEMs. (1) effective contact detection between convex and non-convex particles; (2) relatively easy code implementation as collision is handled by simple sphere-sphere and sphere-wall contact problems; (3) accurate particle surface and volumetric information can be accessed whenever they are required for the inter-phase interactions, which is particularly important for efficient and accurate couplings.

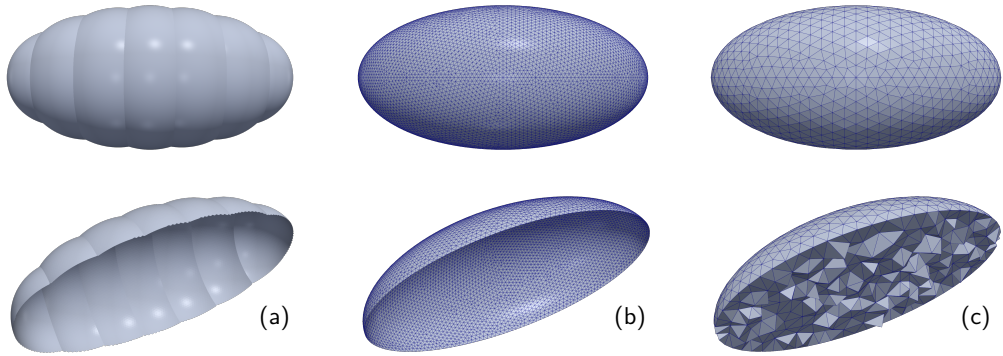

**Figure 3:** Three geometric models and their cut-views for an ellipsoidal particle. (a) MS model with 7 primary spheres; (b) surface mesh model with ca. 10 thousand vertices and 20 thousand triangles; (c) volumetric cell model with ca. 2.5 thousand grid points and 11 thousand tetrahedral cells.

As these extra geometric models are not included in DEM simulation, and only updated at each coupling interval according to the position and orientation of their associated MS model. In the case of ellipsoidal particle shown in Figure 2(b) and Figure 3(b-c), we can use its 15-sphere MS model for very efficient DEM simulations with slightly less accuracy in shape approximation, while accurate surface and cell models for the inter-phase interactions. In typical polyhedral DEMs, if we increase the number of surface elements for a desired resolution by a factor of $n$ ($n > 1$), the complexity of finding closest surface elements between two polyhedral particles is proportional to $O(n^2)$, thus it will become too expensive to use surface mesh of high resolution for polyhedral DEM simulations even with cluster computers.

## 2.2 Neighbor list

Before computing the contact forces acting on a particle, first we need to find its neighbors that are potentially in contact with this particle, because it is not necessary to perform complex and expensive geometry intersection test if two particles are far away from each other. In order to quickly check if two particles are close enough, each particle is usually attached with a minimum bounding sphere (MBS) or minimum bounding box (MBB) that tightly encapsulates the particle [8], as depicted in the 2D schematic in Figure 4. If the minimum bounding volumes (MBS or MBB) of two particles do not intersect, they are considered to be "far away" from each other. This process is usually termed broad phase contact detection, for the purpose of exclusion of particles that are not possible to collide with particle $i$. Since a particle's MBB is usually smaller than the MBS in volume, thus two particles $i$ and $j$ may become neighbors as shown in Figure 4(a) with MBS, while on the opposite with MBB in Figure 4(b). In this work MBS is used for an efficient broad phase process. Nevertheless, if the aspect ratio of particles in granular system is small (e.g. 1:3~1:10), then using MBB can be more efficient for DEM simulations, as there will be fewer neighbors detected in broad phase, even though the MBB intersection test is much more expensive than MBS.

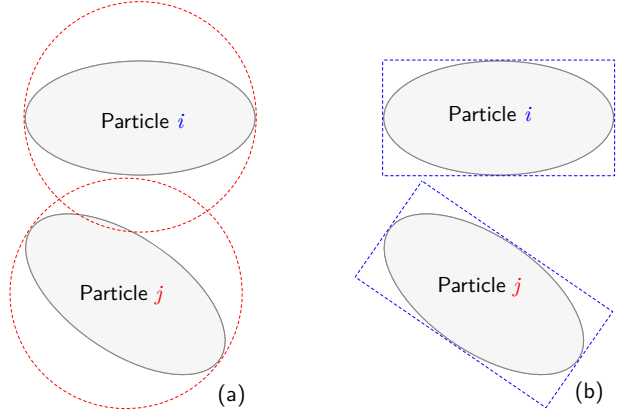

**Figure 4:** 2D schematic of minimum bounding sphere (MBS) and box (MBB) for broad phase contact detection. (a) Particle i and j are neighbors as their MBS are in contact; (b) Particle i and j are not neighbors as their MBB are not in contact.

Finite or infinite walls act as boundaries in DEM simulations. For infinite boundaries such as planar wall that can be defined by $(\mathbf{p}, \hat{\mathbf{n}})$ where $\mathbf{p}$ is a point on the planar wall, $\hat{\mathbf{n}}$ is unit normal vector; or cylindrical wall defined by $(r, \mathbf{p}_1, \mathbf{p}_2)$, here $r$ is the radius, and $\mathbf{p}_1, \mathbf{p}_2$ are two points on the axis of the cylinder, a direct MBS-wall intersection test is performed to check if particle is close to the wall. For finite and complex boundaries triangle mesh is often used, therefore we need to loop over each triangular face of the mesh to particles, which is expensive as a complex mesh wall usually consists of few thousand of triangles. Thus each triangle is encompassed with a MBS and treated as a particle in broad phase.

Considering a granular system with m individual particles and n finite walls (e.g. triangular faces), a trivial broad phase intersection tests for particle-particle and particle-wall contacts is achieved by looping each particle over all other particles and walls, the time to examine all pair separations is proportional to $\frac{1}{2}N(N-1)$ where $N = m + n$, i.e. of order $\mathcal{O}(N^2)$. To avoid this expensive distance evaluation, several strategies have been proposed, namely Verlet list, cell list, and the combination of both. The key concept of Verlet list approach [39] is to add an artificial layer around particle's MBS as "buffer zone" shown in Figure 5(a). The thickness of this layer is termed skin distance. Therefore the radius of particle's bounding sphere $R_m$ reads as

$$R_m = R_{MBS} + R_{cut} + d_{skin} \qquad (1)$$

here $R_{MBS}$ is the radius of MBS, $R_{cut}$ is the cut-off distance for long-range forces (i.e. magnetic or electric fields) and $d_{skin}$ is the skin distance. To check one pair of particles with global index i and j (i < j), the radius of particle i's bounding sphere will be $R_m$ while particle j's will be its MBS. In this way, the expensive $\mathcal{O}(N^2)$ operation for finding particle i's neighbors can be done at certain intervals, as long as no particles can penetrate through the skin region into the force cut-off sphere of particle i, if they are not in the neighbor list of particle $i$. Suppose the interval to update neighbor list is $n$, and fixed DEM time-step is $\Delta t$, then $d_{skin}$ can be estimated by: $d_{skin} = n \cdot V_{max} \cdot \Delta t$, where $V_{max}$ is the magnitude of maximum particle velocity.

The Verlet list algorithm has been proven to be efficient when the number of particle in a granular system is relatively small and the velocities of particles are low [42], since the time of constructing neighbor list for each particle still scales $O(N^2)$. If particles move with high speed, the neighbor list must be updated more frequently, or the skin distance must increase. Both of them will increase the simulation time dramatically.

Unlike Verlet list approach, the cell list algorithm [1] does not evaluate distance between particles. Rather, the computational domain is decomposed into cells as shown in Figure 5(b). Particles are sorted into these cells where their centers of mass are contained, and for each cell, a list of particle index is built and updated at certain interval. The cell size should be at least larger than the maximum diameter of particle's MBS. Therefore, searching the neighbors for a particle is a rapid process: it is only necessary to look at particles in the same cell where the query particle is located (green) and in nearest neighbor cells (gray, 8 adjacent neighbor cells in 2D and 26 in 3D). The advantage of cell list algorithm is that the construction of cell list, i.e. assigning each particle to appropriate cells, scales $O(N)$, thus it is suitable for large scale problem. Depending on the cell size, the number of neighbors to a query particle could vary from few tens to few thousands, which may lead to poor performance in the contact force calculation.

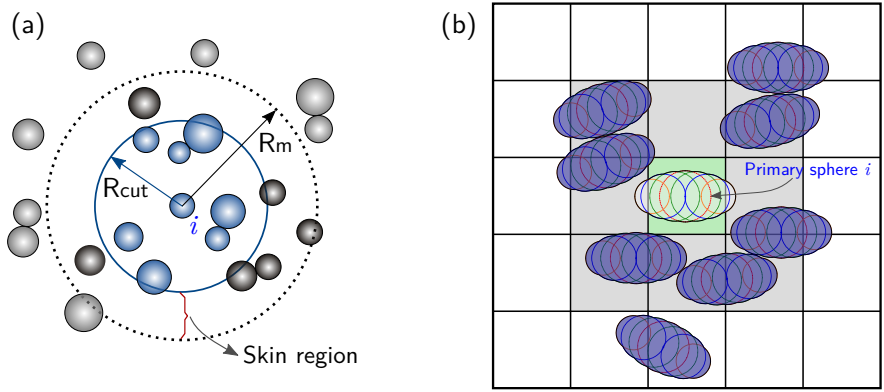

**Figure 5:** Neighbor list algorithms. (a) Verlet list algorithm; (b) Cell list algorithm. Hybrid approach using both Verlet and cell list algorithms is adopted in this work.

It is clear that both Verlet and cell list algorithms have their pros and cons. A more efficient way to build neighbor list with minimum particles is to combine these two into a hybrid approach [e.g. 32]. For every time interval, the cell list is updated, then the search for particle $i$'s neighbors with Verlet algorithm, is narrowed to those particles whose centers are contained in the nearest cells of particle $i$. Since particle's DEM model is made of overlapping spheres in this work, the contact force calculation between two MS particles still scales $O(n^2)$ where $n$ is the number of spheres per particle. For more efficient DEM simulations, a list of primary spheres (may be from different MS particles) is built and updated every time interval for each cell as shown in 5(b), then the Verlet algorithm is used to find a primary sphere $i$'s neighboring spheres in its adjacent cells, but those who belong to the same MS particle will be ignored, as relative positions between primary spheres in a MS particles are fixed in body frame. In this case, the cell size is usually set as the maximum sphere diameter in system by a factor $k$ ($k \in [1, 3]$ in this work), i.e. $d_{cell} = k \cdot D_{max}$. As a result, a primary sphere's neighbors can be limited up to a few tens, thus the contact force calculation for a MS particle is of order $O(m \cdot n)$, where $m$ is the average sphere neighbors, and normally smaller than 10 in most case, for small cell size in neighbor list building.

## 2.3 Contact detection

Once the neighbor lists are built or updated for all MS particle's primary spheres, the complex contact detection between particle-particle and particle-wall is then converted into relatively simple "sphere-to-wall" and "sphere-to-sphere" contact problems. As shown in Figure 6(a), a MS particle (yellow color) is in contact with a flat wall and another particle (gray color). The actual contacts can be resolved by the intersections between the primary sphere $i$ from yellow particle to wall and the primary sphere $j$ from gray particle. This process is usually termed as narrow phase contact detection, as we need to compute the contact points, overlap distance to obtain contact force and torque between two colliding objects.

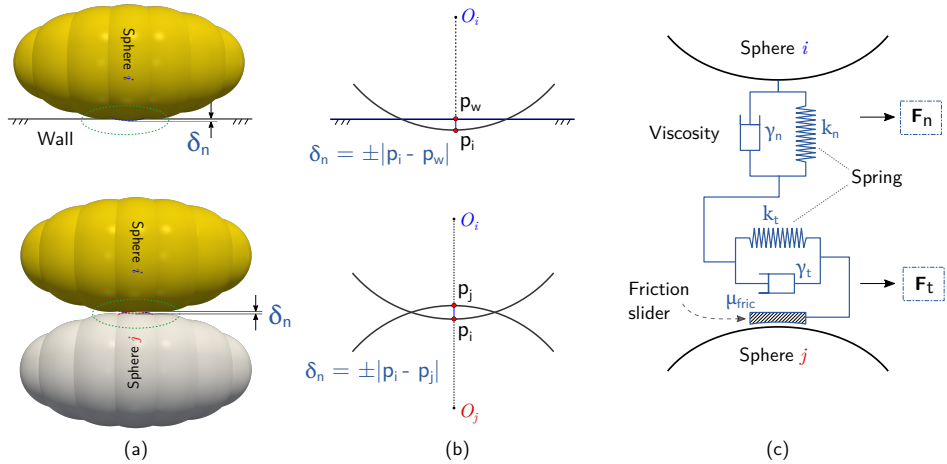

**Figure 6:** Narrow phase contact detection and force model. (a) particle-wall and particle-particle contacts; (b) closer views at contact points; (c) contact model using spring, dash-pot and friction slider.

Sphere-to-wall contact resolution is straightforward, if only planar wall is considered as shown in Figure 6(b). First the sphere center $O_i$ is projected onto the planar wall to obtained the contact point $p_w$. Let point $p_i$ be the intersection between the sphere and a ray (from $O_i$ to $p_w$), then it will be the deepest point on sphere $i$ penetrating into wall. Thus the overlap distance $d_n$ reads as: $\pm|p_i - p_w|$, negative if sphere-wall intersects. In practical implementation, $d_n$ and contact normal $\mathbf{n}$ can be simply computed as follows.

$$d_n = |\mathbf{O}_i - \mathbf{p}_w| - R_i \tag{2}$$

$$\mathbf{n} = \frac{\mathbf{O}_i - \mathbf{p}_w}{|\mathbf{O}_i - \mathbf{p}_w|} \tag{3}$$

For sphere to triangular wall contact, there are three types of intersection tests, namely sphere-plane, sphere-edge and sphere-vertex. Since they are basic geometry problems, the detail will not be discussed here.

As simple sphere-to-sphere contact is responsible for the complex particle-particle contact problem, thus the narrow phase process can be very efficient in MS-DEM, depending on the number of primary spheres per MS particle. As shown in Figure 6(b), the contact points $\mathbf{p}_i$ and $\mathbf{p}_j$ are the intersection between line through sphere centers $\mathbf{O}_i$ and $\mathbf{O}_j$ and two primary spheres i and j, therefore the overlap distance $d_n$ reads as: $\pm|\mathbf{p}_i - \mathbf{p}_j|$. In practical implementation, $d_n$, contact normal $\mathbf{n}$, and contact points $\mathbf{p}_i, \mathbf{p}_j$ can be computed as follows.

$$d_n = |\mathbf{O}_i - \mathbf{O}_j| - (R_i + R_j) \tag{4}$$

$$\mathbf{n} = \frac{\mathbf{O}_i - \mathbf{O}_j}{|\mathbf{O}_i - \mathbf{O}_j|} \tag{5}$$

$$\mathbf{p}_j \approx \mathbf{p}_i = \mathbf{O}_i - R_i \mathbf{n} \quad \text{(for torque calculation)} \tag{6}$$

Note that for torque calculation only one of the contact points is needed, as the distance from $\mathbf{p}_i$ to $\mathbf{p}_j$ is so small ($d_n$ of order $10^{-6}$ m) compared to finite-size sphere radius (of order $10^{-3}$ m).

## 2.4 Contact force and torque

Once the contact information such as overlap distance $d_n$, contact point $\mathbf{p}_i$ and contact normal $\mathbf{n}$ between sphere-wall and sphere-sphere is obtained from the narrow phase process, the pair-wise contact force and torque between particle-particle and particle-wall can be computed with the classic contact model using spring, dash-pot and slider [6] as illustrated in Figure 6(c).

The normal contact force $\mathbf{F}_{n,ij}$ consists of two components: the spring force represents elastic repulsion, while the damping force accounts for the energy dissipation during the collision, and can be written as follows.

$$\mathbf{F}_{n,ij} = k_n \boldsymbol{\delta}_{n,ij} - \gamma_n \mathbf{V}_n \tag{7}$$

where $k_n$ and $\gamma_n$ are the elastic and viscoelastic damping constants for normal contact. $\boldsymbol{\delta}_{n,ij}$ is the overlap distance vector ($d_n \mathbf{n}$), and $\mathbf{V}_n$ is the normal component of the relative velocity $V_{rel}$ of two particles or particle-wall at the contact point. In general, there are two types of model for the calculation of $k_n$, namely Hooke (linear) model, and Hertz (non-linear) model [35]. As pointed out by Zheng et al. [47], the linear model is not as accurate as other semi-theoretical models (e.g. Hertz model), and has obvious disadvantage in determining the contact area and duration time, thus the non-linear Hertz-Mindlin (HM) model [35] is adopted here to compute normal and tangential coefficients.

$$k_n = \frac{4}{3} Y^* \sqrt{R^* |d_n|} \tag{8}$$

$$\gamma_n = \sqrt{5} |\beta| \sqrt{m^* k_n} \tag{9}$$

here $Y^*$, $R^*$ and $m^*$ are the effective Young's modulus, radius and mass, respectively. $\beta$ is a function of the restitution coefficient $e$. More details on these coefficients one can refer to SR-DEM.

Similarly, the tangential contact force $F_{t,ij}$, which also consists of elastic and damping components, can be written in the following form.

$$F_{t,ij} = k_t \delta_{t,ij} - \gamma_t V_t \tag{10}$$

here $k_t$ and $\gamma_t$ are the elastic and viscoelastic damping constants for tangential contact. $\delta_{t,ij}$ is the accumulated tangential displacement vector, i.e. contact history, reads as

$$\delta_{t,ij} = -\int_{t_0}^{t_1} V_t(\tau) d\tau \tag{11}$$

here $t_0$ and $t_1$ are the begin and end of the contact, thus the contact duration is $t_1 - t_0$. In order to account for the change of contact normal, i.e. the relative position of the contacting pair at contact point, the old $\delta_{t,ij}$ computed from last DEM time-step needs to be rotated to the new tangential plane [26] as follows.

$$\delta_{t,ij}^{(n)} = \left[\delta_{t,ij} - (\delta_{t,ij} \cdot n)n\right] \frac{|\delta_{t,ij}|}{\left|\delta_{t,ij} - (\delta_{t,ij} \cdot n)n\right|} \tag{12}$$

here $n$ is the new contact normal. Therefore the accumulated tangential displacement vector at time-step $n + 1$ can be written as

$$\delta_{t,ij}^{(n+1)} = \delta_{t,ij}^{(n)} - V_t^{(n+1)} \Delta t \tag{13}$$

For efficiency, a simple relationship for $k_t$ and $\gamma_t$ is employed: $k_t/k_n = 2/7$ and $\gamma_t/\gamma_n = 1/2$, as the contact dynamics are not very sensitive to precise values of these ratios [37].

The tangential force will be truncated if it exceeds the Coulomb friction, because the slip between a contacting pair occurs in this case. Thus the tangential force $F_{t,ij}$ has the following form.

$$F_{t,ij} = \begin{cases} k_t \delta_{t,ij} - \gamma_t V_t & \text{if } |F_{t,ij}| < \mu_{fric}|F_{n,ij}| \\ -\mu_{fric}|F_{n,ij}|\frac{V_t}{|V_t|} & \text{if } |F_{t,ij}| \geq \mu_{fric}|F_{n,ij}| \end{cases} \quad (14)$$

where $\mu_{fric}$ is the friction coefficient. The relative velocity $V_{rel}$ of two contacting particles $i$ and $j$ at the contact point can be written as

$$V_{rel} = V_{i,cp} - V_{j,cp} = [V_i + \omega_i \times (p_c - C_i)] - [V_j + \omega_j \times (p_c - C_j)] \quad (15)$$

here $p_c$ is the contact point, $C_i$ and $C_j$ are the center of mass for particle $i$ and particle $j$, respectively. $V_i$, $V_j$ and $\omega_i$, $\omega_j$ are the translational and rotational speed stored at the particle center. Therefore the normal and tangential components of the relative velocity can be calculated as follows.

$$V_n = (V_{rel} \cdot n)n \quad (16)$$
$$V_t = V_{rel} - V_n \quad (17)$$

Finally, the contact force acting on particle $i$ can be assembled $F_{ij} = F_{n,ij} + F_{t,ij}$. Calculation of torques (i.e. rotational forces) acting on particles is straightforward.

$$\text{Torque on particle-}i: \quad T_{ij} = (p_c - C_i) \times F_{ij} \quad (18)$$
$$\text{Torque on particle-}j: \quad T_{ji} = (p_c - C_j) \times (-F_{ij}) \quad (19)$$

## 2.5 Time integration

Once the contact forces and torques acting on a particle $i$ from all other particles and walls are computed, its transnational and rotational velocities, position and orientation can be therefore updated within a small time duration. Equations of motion by applying Newton's second law read as

$$m_i \frac{dV_i}{dt} = \sum_{j=1}^{N_c} (F_{n,ij} + F_{t,ij}) + F_b \quad (20)$$

$$I_i \frac{d\omega_i}{dt} = \sum_{j=1}^{N_c} (T_{ij} + T_{r,ij}) \quad (21)$$

where $N_c$ is the number of contacts to particle-$i$, $F_b$ is body force (e.g. gravity) acting on particle $i$. Rolling friction torque $T_{r,ij}$ is neglected in this work, as rolling friction is usually of several orders of magnitude smaller than sliding friction for particles with relatively smooth surface. As particle's inertia tensor $I_i$ in world frame is not constant, Eqn. (21) is often computed in particle's body frame for simplicity.

Second-order Velocity-Verlet algorithm [38] is adopted in this work to integrate the equations of motion (20-21), among other time integration schemes reviewed and compared in literature [7, 18, 36], due to its efficiency, low complexity and acceptable accuracy.

$$\mathbf{X}(t + \Delta t) = \mathbf{X}(t) + \mathbf{V}(t)\Delta t + \frac{1}{2}\mathbf{a}(t)\Delta t^2 \qquad (22)$$

$$\mathbf{V}(t + \Delta t) = \mathbf{V}(t) + \frac{1}{2}\big[\mathbf{a}(t) + \mathbf{a}(t + \Delta t)\big]\Delta t \qquad (23)$$

where $\mathbf{X}(t + \Delta t)$, $\mathbf{V}(t + \Delta t)$ are the updated particle position vector and velocity at new time $t + \Delta t$ from known velocity, position and net force at old time t.

An appropriate time-step $\Delta t$ in the time integration is important for a stable and efficient DEM simulation. There are several models to estimate two-particle collision time or so-called critical time step ($\Delta t_c$), which is similar to the Courant number in computational fluid dynamics (CFD) to ensure the numerical stability. In this work Rayleigh time $T_R$ based on Rayleigh wave speed [30], and Hertz time $T_H$ based on Hertz theory [15] are employed for the critical time step estimation.

$$\Delta t_c = \min\{T_R, T_H\}, \text{ where } T_R = \frac{\pi \overline{R}}{K}\sqrt{\frac{\rho}{G}}, \ T_H = 2.8683\left(\frac{m^{*2}}{R^* \, Y^{*2} \, \mathbf{V}}\right)^{0.2} \qquad (24)$$

here $Y^*$, $R^*$ and $m^*$ are the same coefficients in Eqn. (9). Coefficient K is a function of Poisson's ratio $\nu$, reads $0.8766 + 0.1631\nu$. $\rho$ and G are the particle's density and shear modulus, respectively. For a more conservative estimation, average particle radius $\overline{R}$ is set as minimum particle radius $R_{min}$, and particle's velocity $\mathbf{V}$ is the maximum in system. Finally $\Delta t = \Delta t_c/n$, where n is usually set in the range between 10 to 100 to further ensure the numerical stability in DEM simulations.

## 3. Validation

Four different cases were performed to quantitatively verify and validate the hybrid DEM framework Rigid3D using the multi-sphere approach as the engine of DEM simulations. Particle-wall and particle-particle contact pairs are tested in the first case, in order to verify the non-linear Hertz-Mindlin model described in section 2.4. In the second case static packing of particles with 5 different shapes in rectangular and cylindrical containers is carried out. The third case uses the packed particles from the second validation case to perform the particle slumping test under gravity. Ellipsoidal particles tumbling in a cylindrical, rotating drum is simulated in the fourth case, and compared with experimental results to investigate the mixture behavior. Note that the particle shapes in all cases are relatively simple, so that we can construct accurate MS models without involving complex algorithms designed for arbitrary shapes [e.g. 44].

## 3.1 Particle-wall and particle-particle contact

This case aims to verify the contact detection algorithm, and the effectiveness of non-linear Hertz-Mindlin contact model in Rigid3D. Particle and wall are assumed to be frictionless and gravity is neglected. An ellipsoidal particle, initially set a translational velocity $(0, 0, -1)$ m/s and zero angular speed, impacts a fixed horizontal wall, and another identical and fixed ellipsoidal particle shown in Figure 7-8. Two particles are parallel to each other and their centers of mass are aligned in the $z$-direction. The material properties and simulation parameters are listed in Table 1.

**Table 1:** Parameters used in particle to wall/particle impact simulation

| Parameter | Value |
|---|---|
| Ellipsoid semi-axes, a, b, c (mm) | 5, 2.5, 2.5 |
| Young's modulus (GPa) | 10.0 |
| Poisson's ratio | 0.3 |
| Density (kg/m$^3$) | 2500.0 |
| Coefficient of restitution | 0.6 |
| Time step $\Delta t$ (s) | $1.0 \times 10^{-7}$ |

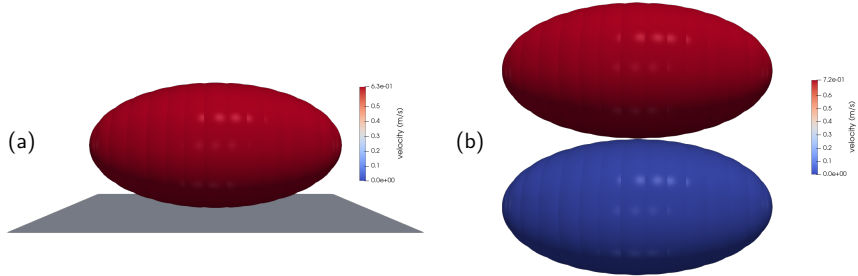

**Figure 7:** Velocity profile a moment after the head-on contacts between (a) particle-wall and (b) two identical ellipsoidal particles.

In the case of particle-wall head-on contact shown in Figure 7(a), the contact normal from the contact point passes through the particle's center of mass, thus the momentum arm is zero. The dimensionless rebound velocity 0.63 simulated in Rigid3D is close to the prescribed coefficient of restitution 0.6 with a small error of 5%. If the Young's modulus is reduced to 1.06 Gpa, then the rebound velocity will match the coefficient of restitution perfectly. For the particle-particle head-on contact in Figure 7(b), the rebound velocity is constant with value of 0.72, if the Young's modulus is no less than 0.5 Gpa. The reason why the dimensionless rebound velocity is greater than 0.6, is that the normal damping coefficient in particle-particle contact is smaller than that of in particle-wall contact, thus less energy dissipation. According to the Eqn. (9), $\gamma_n^{(pp)}/\gamma_n^{(pw)}$ will be $0.5^{1/8} \approx 0.92$ at the same overlap distance, where superscripts $pp$ and $pw$ denote the particle-particle and particle-wall contact. From the contact mechanism point of view, the contact area at the contact point between two convex surfaces is smaller than that of between convex and flat surfaces for the same deformation, therefore less energy is dissipated during the compression and restitution phases.

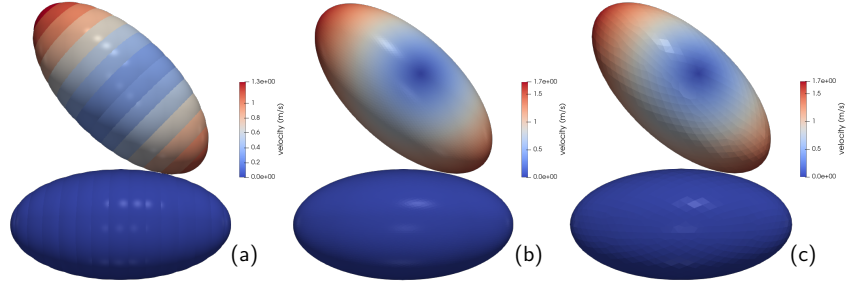

**Figure 8:** Velocity profile of the elements from (a) MS model, (b) surface model and (c) cell model a moment after the oblique contact (45 degree) between two ellipsoidal particles.

The post-impact angular velocity of the particle (top) in the oblique contact is non-zero in Figure 8, thus the velocity profile on the particle surface is not uniform. Since a sphere is implicitly defined by its center and radius, it is difficult to access particle's surface and volumetric information through the MS model. As depicted in Figure 8(a), primary spheres are colored by the velocity magnitude at sphere center via $\boldsymbol{V_c} + \boldsymbol{w} \times \boldsymbol{r_i}$, here $\boldsymbol{V_c}$, $\boldsymbol{w}$ are particle's translational velocity at center of mass and angular velocity respectively, $\boldsymbol{r_i}$ is the position vector from particle's center to sphere-i's center. In case of inter-phase coupling, the accurate surface and volumetric information can be accessed by particle's surface model (Figure 8b) and cell model (Figure 8c).

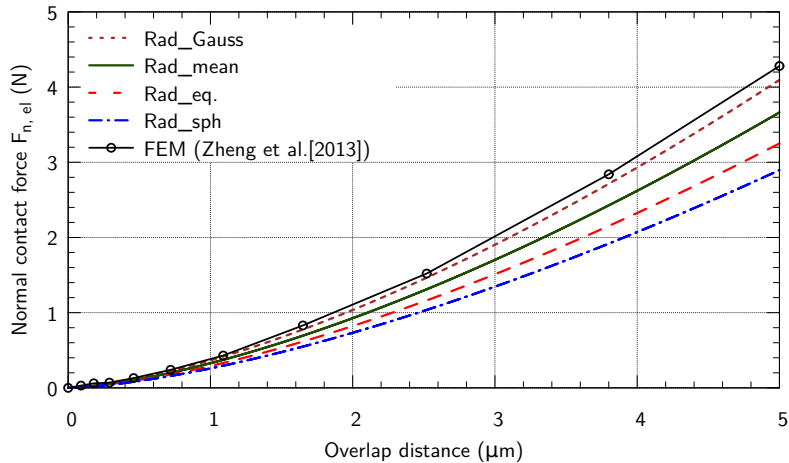

**Figure 9:** Normal contact force vs. overlap distance in head-on and oblique contacts.

The elastic component $F_{n,el} = |k_n \delta_{n,ij}|$ of the normal contact force in the two particles head-on contact, is plotted in Figure 9 as a function of the overlap magnitude in the range [0, 5] µm, and compared with the FEM analysis carried out by Zheng et al. [47]. Four radius of curvature models are tested: the Gaussian and mean curvature radii ($R_G$, $R_m$) from the surface model, radius of the particle's equivalent volume sphere ($R_{eq}$), and radius of MS model's primary sphere at contact point ($R_{sph}$). Figure 9 shows that the Gaussian curvature model is able to reproduce the FEM results most closely among all models by the Hertz theory. As the elastic stiffness $k_n$ is proportional to the square root of effective radius, whose value at the contact point is: $R_G^* > R_m^* > R_{eq}^* > R_{sph}^*$, therefore the elastic normal force $F_{n,el}$ is also proportional to the square root of the radii of curvature.

Despite the difference in the normal contact forces from various radius of curvature models, the dimensionless rebound velocities at the center of mass still converge in all contact types (Figure 7-8) as shown in Figure 10(a). It is because the ratio of calculated normal elastic stiffness $k_n$ between different radius of curvature models is in the range $[1, \sqrt{2}]$. Since the rebound velocity is not sensitive about the Young's modulus thus the normal elastic stiffness. It implies that we do not have to do the expensive calculation of the Gaussian or mean curvature at the contact point from particle's surface model. Nevertheless, smaller elastic stiffness (softer) does lead to slightly longer contact duration as shown in Figure 10(b), which might cause small deviation in particle's dynamics. It was tested if the Young's modulus is no less that 50 MPa, the different between the radius of curvature models is negligible. Therefore in this work, only the $R_{sph}$ and $R_{eq}$ radius of curvature models are used because there is no additional cost. It is suggested to use $R_{sph}$ model if particle has sharp corners in order to reflect real curvatures, while the $R_{eq}$ model is suitable for relatively rounded particles.

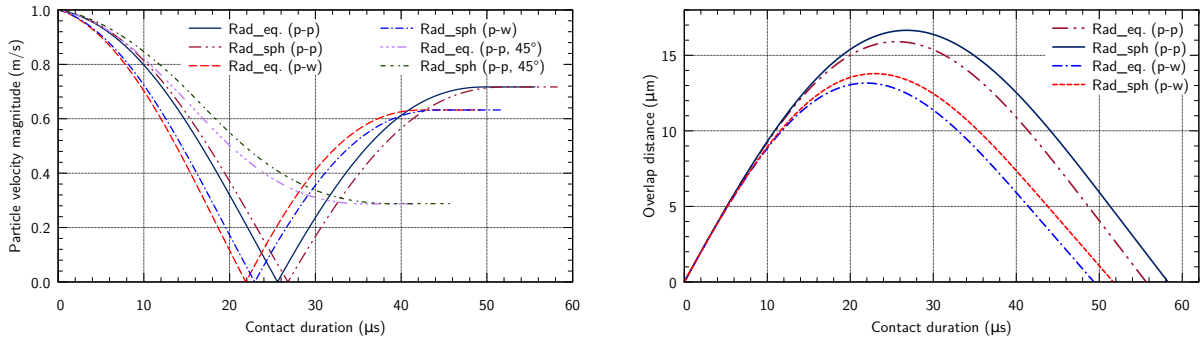

**Figure 10:** (a) dimensionless velocity at the center of mass, and (b) overlap distance as a function of the contact duration in head-on and oblique contacts.

## 3.2 Particle packing in containers

Packing of particles is a common process in pharmaceutical, chemical, agricultural and food industries, for minimization of the occupation space and maximization of the packing strength, etc. Particle shape has been identified as one of the most important particle properties that can affect the packing structure, among other factors such as packing method and container conditions [12]. This validation consists of two parts: (1) comparison of the experimental [24] and the DEM simulation fill heights and packing porosity of capsules in a cylindrical container; (2) numerical study of the particle shape effect on the particle packing porosity in a rectangular container.

The wall of right circular cylinder in this work is implicit, defined by the cylinder axis and radius, so that the contact resolution between sphere and cylindrical wall can be very efficient and accurate, because an acceptable approximation of mesh based cylindrical wall usually requires few thousand triangles. Therefore an open cylindrical container in the DEM simulation can be described by an implicit cylindrical wall and a flat wall as the bottom. In the following case, the internal diameter and height of the container are 93.5 mm and 195 mm, respectively. A total of 600 capsules are dropped from the top of the container with zero velocity, random orientation and lateral positions. In the DEM simulation, a particle stream at the same location is created in order to mimic the particle packing process (see e.g. Figure 13): at certain time interval 6-10 capsules are generated from the top of the container, and free fall under gravity onto the bottom of the container. The additional physical and numerical parameters of the capsule packing simulation are listed in Table 2.

**Table 2:** Parameters used in the capsules packing simulation

| Parameter | Value |
| --- | --- |
| Capsule diameter (mm) | 7.6 |
| Capsule height (mm) | 21.4 |
| Density (kg/m$^3$) | 917.0 |
| Number of particles | 600 |
| Cylinder diameter (mm) | 93.5 |
| Cylinder height (mm) | 195.0 |
| Young's modulus (Pa) | $5.0 \times 10^7$ |
| Poisson's ratio | 0.3 |
| Coefficient of friction | $0.3^{(pw)}$, $0.4^{(pp)}$ |
| Coefficient of restitution | 0.6 |
| Time step $\Delta t$ (s) | $5.0 \times 10^{-7}$ |

Superscripts $pw$ and $pp$ denote particle-wall and particle-particle coefficients.

In order to obtain the final fill height of capsules from the DEM simulation, screenshots of two perpendicular side views (e.g. Figure 11a) of the cylindrical container were taken to evaluate the mean fill height. In this work a simple yet efficient image process approach is adopted: first a spline that closely fits the free surface of the packed capsules is drawn, then the spline is subdivided into N (N > 100) line segments of equal length, while normally it takes only 20 to 50 actual sampling points to form an accurate spline of the (2D) packing free surface. These N + 1 sampling points will be used to calculate the mean fill height. The final mean will be the average of the mean fill heights from all side views.

The final packing states between the simulation and experiment is compared in Figure 11. The final fill height in the experiment is 122.9 ± 4.0 mm [24], while the calculated value in the Rigid3D DEM simulation is 126.6 ± 3.5 mm. The error of the fill heights is about 3%, which is in good agreement with the experiment. The reason why the fill height in simulation is slightly over-predicted, is that the real capsule is composed of two halves: a smaller-diameter "body" that is filled and then sealed using a larger-diameter "cap", while the capsule's MS model (21 primary spheres) in the simulation consists of three parts: two semi-spheres and one cylinder with the larger diameter of the real capsule's cap. The packing porosity is then calculated as $1 - \sum_{i=1}^{n} V_i/Ah$, where $n$ is the number of capsules, $V_i$ is the volume of $i^{th}$ capsule, $h$ is the fill height and $A$ is the area of the container bottom base.

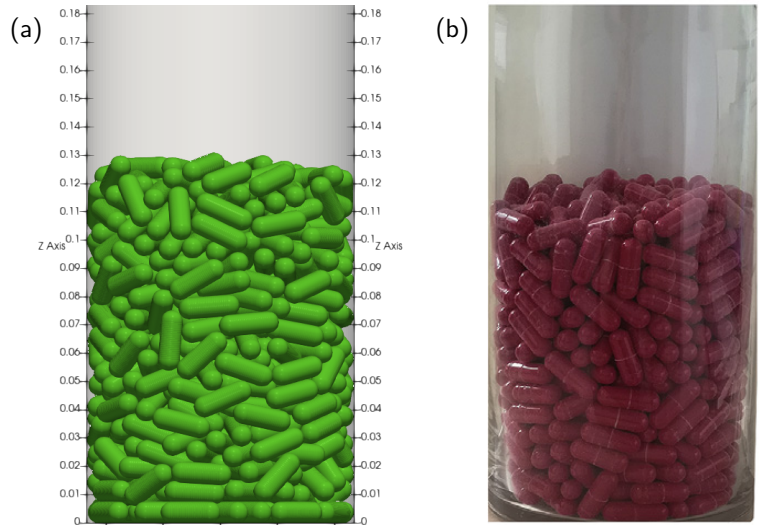

**Figure 11:** Packing of 600 capsules in a cylindrical container. (a) final state in Rigid3D simulation; (b) experimental result [24].

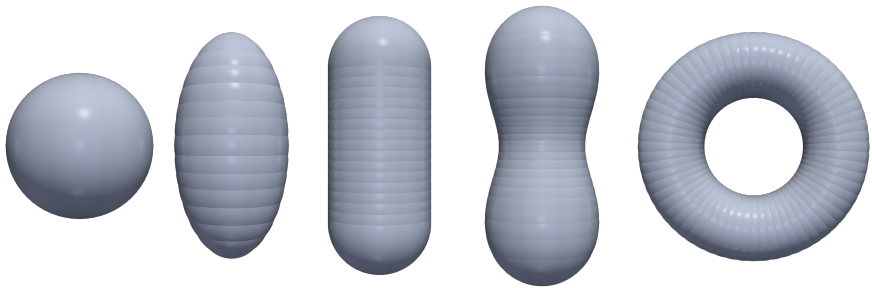

**Figure 12:** 5 multi-sphere models of same volume from left to right: sphere, ellipsoid (15 spheres), sphero-cylinder (17 spheres), 3D Cassini oval (29 spheres) and torus (64 spheres).

It seems that the Rigid3D code is able to reproduce the particle packing process, thus we can perform more numerical experiments of particle packing using the same simulation set-up, in order to study the effect of particle and container shapes on the packing porosity. In the following sub-case, we examine another 5 particle shapes of same volume shown in Figure 12: a sphere, an ellipsoid, a sphero-cylinder, a torus, and a peanut-shaped particle which is the solid of revolution by rotating a Cassini oval around its major axis (named 3D Cassini oval in this work). The details of the particle shapes are listed in Table 3. For the sphero-cylinder, L is the cylinder length, R is the radius of the cylinder and semi-spheres.

The capsule (or sphero-cylinder) used in the previous validation has slightly larger cylinder length to semi-sphere radius ratio: $L \approx 3.63R$. In the torus implicit equation, R is the major radius, which is the distance from the center of the tube to the center of the torus; while r is the minor radius, which is the radius of the tube. Note that the ellipsoid here is taken from the first validation case, while the other four shapes are scaled to the same volume of the ellipsoid.

**Table 3:** Particle shapes for the numerical experiments

| Particle shape | Parameters |
| --- | --- |
| Sphere: $\frac{x^2}{a^2} + \frac{y^2}{b^2} + \frac{z^2}{c^2} = 1$ | $a = b = c$ |
| Ellipsoid: $\frac{x^2}{a^2} + \frac{y^2}{b^2} + \frac{z^2}{c^2} = 1$ | $a = 2b = 2c = 5$ mm |
| Sphero-cylinder: two semi-spheres connected by a cylinder | $L = 3R$ |
| Cassini oval: $((x^2 - a^2) + y^2)((x^2 + a^2) + y^2) = b^4$ | $b = 1.1a$ |
| Torus: $\left(\sqrt{x^2 + y^2} - R\right)^2 + z^2 = r^2$ | $R = 2.5r$ |

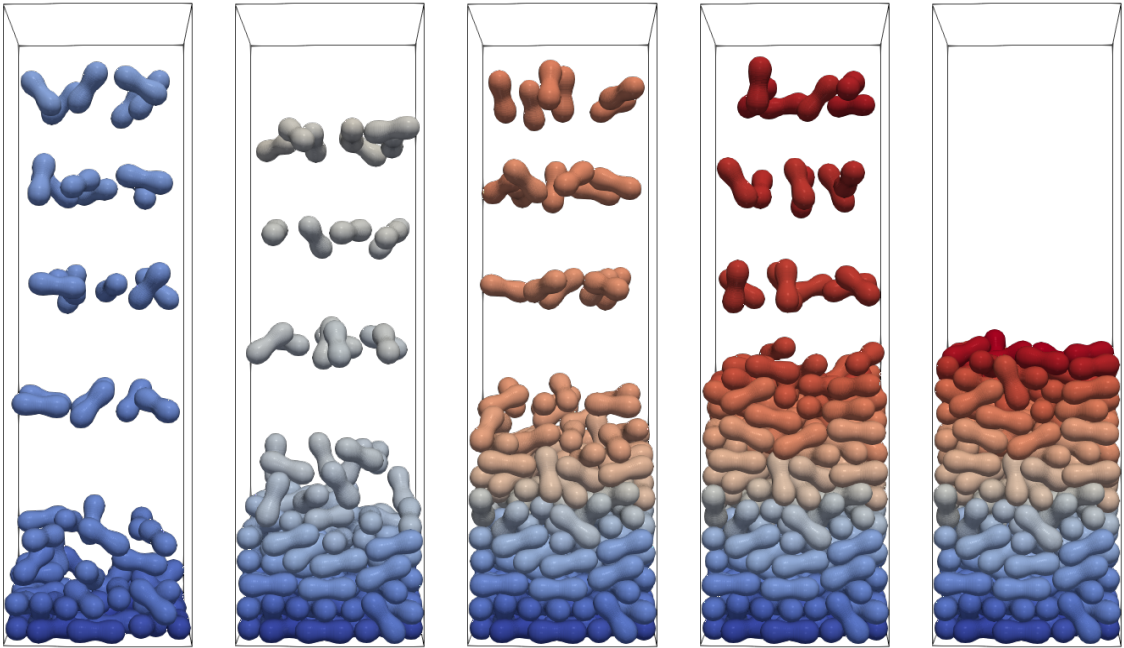

**Figure 13:** Screenshots of the particle (3D Cassini oval) packing process in a rectangular container. Here particles are colored by the global particle index.

For all the shapes, 300 particles are dropped in a rectangular container with a dimension of 35 mm × 35 mm × 120 mm. The particle packing process is simulated by setting a stream of poured particles slightly below the top of the container as shown in Figure 13. The particle stream is implemented in the following steps: (1) a specified region is created for the particle inserting. This region can be simply defined by a box or cylinder. For irregular regions, a triangle mesh is imported as the center of the region, and extruded in the direction and the opposite direction of the faces (i.e. triangles) normal by a length of $h$. The extrusion volume will be $2Ah$ where $A$ is the surface area of the mesh. (2) particles are generated one by one with random center of mass and orientation in the region of particle stream, as long as their minimum bounding spheres do not intersect with walls and previously generated particles. The particle inserting process will stop until a specified particle mass or number is reached, or simply no more particles can be inserted in the stream region. A minimal interval to trigger the particle stream is evaluated before the simulation for efficiency, based on the specified initial particle velocity and the height of the insert region normal to the velocity direction. In the rectangular container packing, the interval of particle stream is set to $10^5$ steps, so that previously generated particles from the stream can more or less settle on the top of the packed particles as shown in Figure 13.

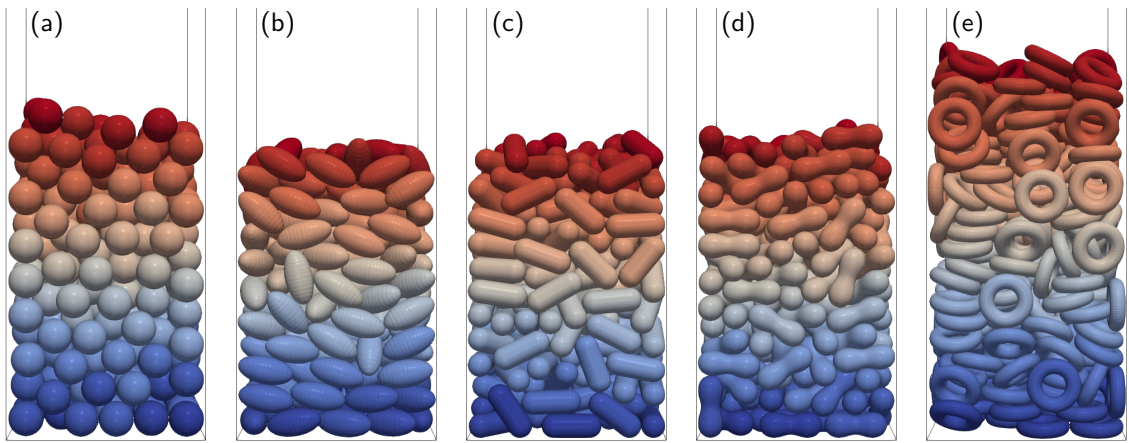

**Figure 14:** Packing of 300 particles (colored by global index) of 5 different shapes in a rectangular container. (a) sphere, (b) ellipsoid, (c) sphero-cylinder, (d) 3D Cassini oval and (e) torus. Note that any particle in the tests has same volume.

**Table 4:** Estimated packing porosity of the particle beds in the rectangular container

| Particle shape | Porosity (%) |
| --- | --- |
| Packing of 600 capsules in a cylindrical container | |
|     Capsule: experiment [24] | 39.1 |
|     Capsule: Rigid3D simulation | 40.9 |
| Packing of 300 particles in a rectangular container | |
|     Sphere | 43.9 |
|     Ellipsoid | 40.6 |
|     Sphero-cylinder | 41.3 |
|     3D Cassini oval | 42.2 |
|     Torus | 54.5 |

The final packing states of the 5 particle packings are presented in Figure 14. Apparently the particle shape has strong impact on the packing porosity, which varies linearly with the fill heights that are evaluated by the same image process used above. The packing porosity of different configurations is computed and listed in Tablet 4. For the torus packing, the porosity (ca. 54.5%) is significantly higher than the rest, as the void space in the middle of torus can not be filled by other toruses. The difference in porosity is marginal in the range from 39.1% to 42.2%, among the capsule packing in the cylindrical container, and the ellipsoid, sphero-cylinder and 3D Cassini oval packings in the rectangular container, despite the particle sizes and container shapes are quite different. The reason for this may be due to the similar particle aspect ratio between 2.0 and 2.81, which is defined as the ratio of the longest edge to the shortest edge of particle's minimum bounding box. The packing of mono-sized spheres exhibits slightly higher porosity (ca. 43.9%) than that of elongated particles of same volume, which is consistent with previous studies [e.g. 27, 41, 48]. There are other important factors that can affect the packing properties, such as packing methods, particle's sliding friction coefficient, dropping height and deposition intensity (e.g. number of particles fed into the container per unit time) [21], however, the numerical study of these factors goes beyond the present paper, as our main goal here is to validate the Rigid3D code, and demonstrate the effect of particle shape on the packing porosity.

## 3.3 Granular dam break

The collapse of a granular column (or dam break) is a classical configuration to understand the flow dynamic of granular material, as the experimental setup is cheap and experiments are easy to conduct [e.g. 2, 19, 34]. This test case will re-use the mono-sized particle packings shown in Figure 14 to perform the numerical simulations of granular dam break in a rectangular channel, with the dimension of 35 mm × 120 mm × 80 mm. The simulation setup consists of a column of packed particles at rest on the left side in the rectangular channel as shown in Figure 15(a). At time t = 0 s, a vertical barrier that holds the particles is removed, and the column of particles start to collapse under gravity onto the flat surface.

Successive snapshots from the simulation of the 3D Cassini oval packing collapse shown in Figure 15 illustrate a typical dry granular dam break of non-spherical particles. Once all particles settle down on the flat surface, the same image process described in section 3.2 can be used to obtain the sampling points of the slope's free surface, from the rear and front views as shown in Figure 16(d). The static angles of repose (AoR) of the rear and front profiles can be estimated by fitting those sampling points with the linear least squares method, thus the final AoR is the averaged value of the two profiles.

The final states of the granular dam break for the 5 different particle shapes are compared in Figure 16. The measured angles of repose of the granular heaps are listed in Table 5. The results show that the particle shape has significant impact on the behavior of granular flow. The angle of repose of the spherical particle heap (Figure 16a) is 8.7°, the smallest as expected, as spherical particles tend to have less inter-particle interlocking than non-spherical particles. Besides, the coefficient of rolling friction in this work is set as 0.001 by default, thus the rolling resistance can be neglected, as it is so small compared to the sliding friction. Since the surface of a particle's multi-sphere model is always concave, thus artificial surface roughness is added and the interlocking between particles is enhanced to some extent.

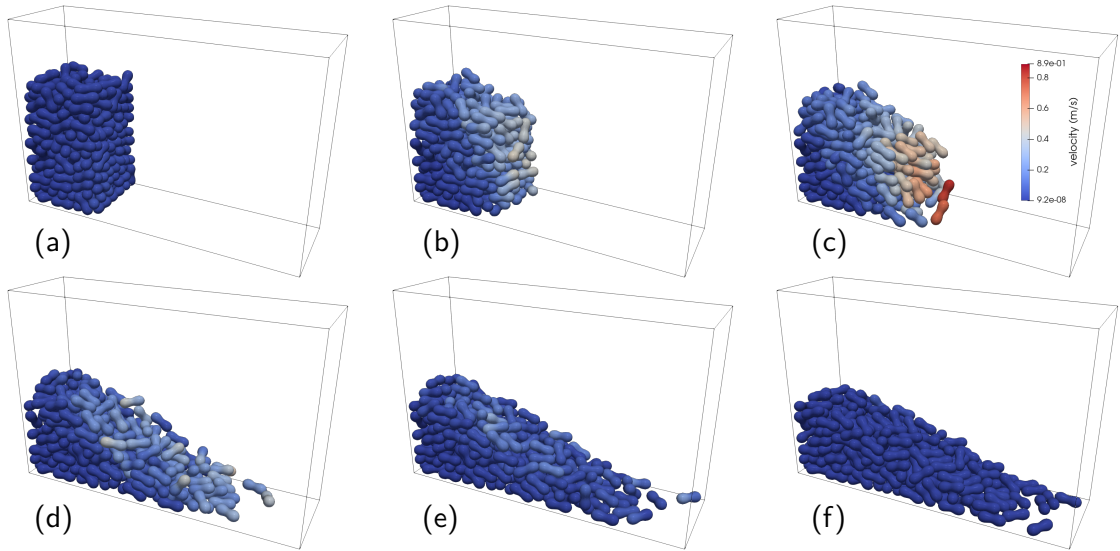

**Figure 15:** Successive snapshots of the collapse of 3D Cassini oval packing in a rectangular channel.

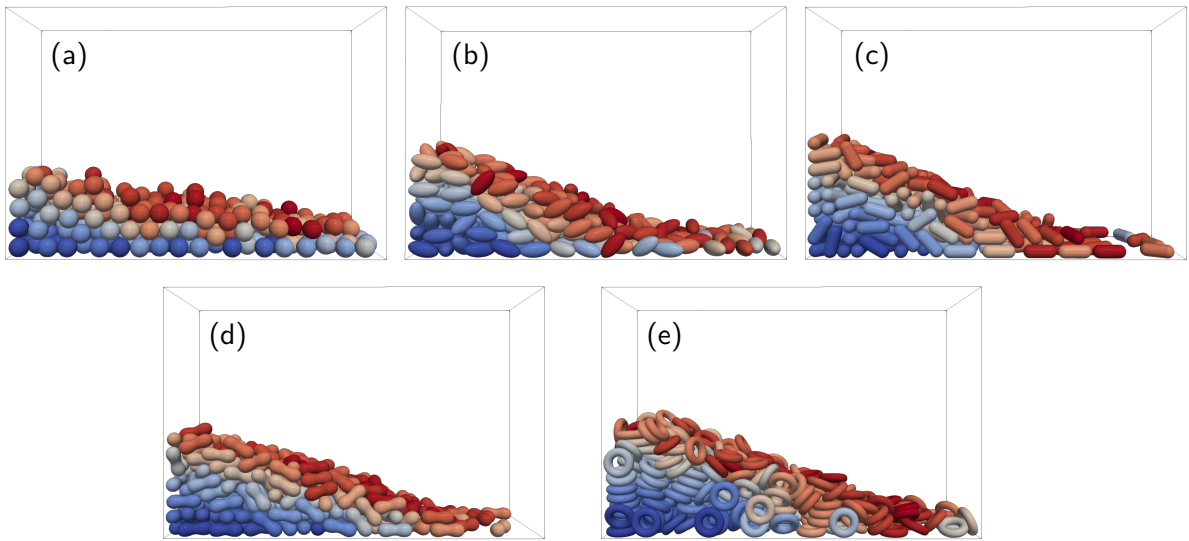

**Figure 16:** Final states of the granular dam break (colored by global index) for the 300 particles of 5 different shapes in a rectangular container. (a) sphere, (b) ellipsoid, (c) sphero-cylinder, (d) 3D Cassini oval and (e) torus.

**Table 5:** Estimated angles of repose for the granular heaps in Figure 16

| Particle shape | Angle of repose (°) |
| --- | --- |
| Sphere | 8.7 |
| Ellipsoid | 21.5 |
| Sphero-cylinder | 24.2 |
| 3D Cassini oval | 18.1 |
| Torus | 22.3 |

For the rest 4 non-spherical particle shapes, the corresponding angles of repose of the granular heaps are significantly larger in the range between 18.1° and 24.2°. The AoR of the 3D Cassini oval (or peanut-shaped) heap is noticeably smaller than the other 3 non-spherical heaps. Possible reasons could be that the surface of the multi-sphere model is smoother (29 primary spheres, see Figure 12), and the particles are further condensed during the granular collapse, as the concave surface allows other particles to fill the gap. The initial height of the granular column can also affect the final angle of repose. As each granular column has the same number (300) of particles of same mass, thus the higher the granular column, the more potential energy it has. Therefore there is more energy to overcome the friction between particle-particle and particle-wall during the granular column collapse, and less resistance to the granular flow. The granular column of torus is about 20% higher than the rest granular columns, thus the AoR (22.3°) could be larger if the column height is about the same to the rest. There are many other factors that could affect the final angle of repose, but in this work we only examine the effect of particle shape.

## 3.4 Rotating drum

The Rigid3D code is further validated in a simulation of particle flow in a rotating drum, which is commonly used for particle coating, drying, mixing or segregation in various industries. In the experiment [43] and DEM simulation, a drum of 200 mm internal diameter and 20 mm thickness, was initially filled with 1000 ellipsoidal particles as shown in Figure 17(a), and tumbled at a rotational speed of 20 rpm. In order to observe the mixing behavior clearly, the particle bed was divided into two layers in equal number of red and blue particles. The size of the ellipsoidal particle and its MS model are identical to the one in Figure 12 (2nd). The material properties and DEM simulation parameters can be found in Table 2, except that the particle density here is 1150 kg/m$^3$.

Two snapshots were taken from the experiment in [43] and the Rigid3D simulation respectively, when the drum rotated 1 revolution and 2 revolutions, and compared in Figure 17(b-c). The flow patterns and the mixture of red and blue particles between our simulation and the experiment have a satisfactory agreement, which further verifies the ability of the code to reproduce granular flow in a dynamic system of non-spherical particles. Even though extensive analysis such as the dynamic angle of repose of the particle bed, Lacey mixing index, and the percentage of the particles reaching the free surface [e.g. 23] will not be carried out in this work, it is quite computationally cheap for the purpose of validation, and we can also gain confidence in the simulated results of similar problems.

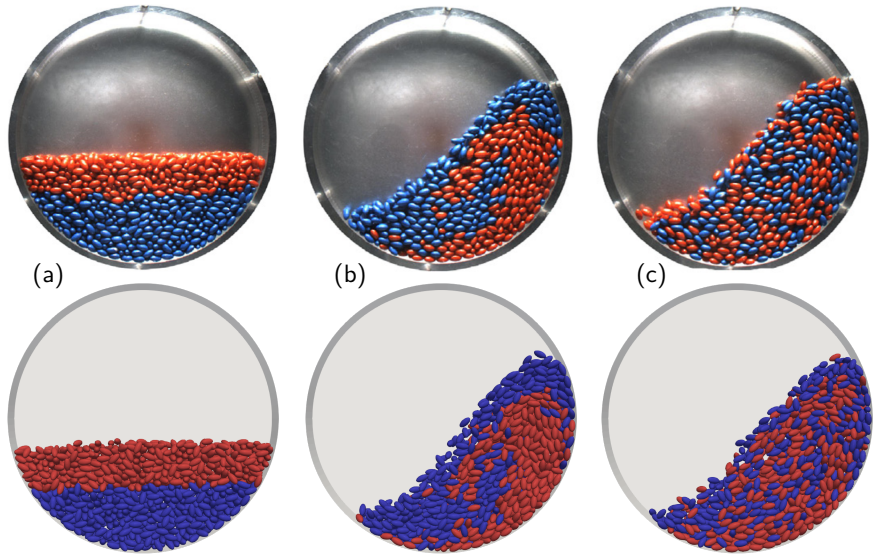

**Figure 17:** Comparison of the experiment [43] and the Rigid3D simulation (bottom) of 1000 ellipsoidal particles in a rotating drum at a speed of 20 rpm. (a) initial status; (b) 1 revolution; (c) 2 revolutions.

## 4. Discussion and perspectives

We suggested a hybrid DEM framework for the purpose of efficient simulation of arbitrarily shaped particles in multi-phase flows. The novelty in this work is to use different particle models for efficient contact detection and the particle related inter-phase interactions, as the details of particle's surface and volumetric information are often needed, e.g. in the process of granular coating, heating transfer and particle-laden flows. In the present work a non-spherical particle is represented by three geometric models: (1) multi-sphere (MS) model for the actual contact resolution in the DEM simulation; (2) surface model, i.e. the particle's surface mesh for accurate coupling with dispersed phases (e.g. droplets); and (3) cell mode, i.e. the discretized particle body for the interaction between particles and continuous phases (gas or liquid). The use of separate particle models enables us to control the balance between the computational efficiency and accuracy for each type of particle related inter-phase interaction.

Since a good approximation of non-spherical particle normally needs tens to thousands of geometric elements (e.g. sphere, triangle or vertex) depending on the shape complexity, and the granular flow is fully resolved in DEM simulations, which is usually the most expensive part, hence the bottleneck of the computational efficiency in a multi-phase flow simulation. We use the multi-sphere model for the contact detection and force calculation in the present work as the "engine" of DEM simulations, because it is relatively easier to implement than other variants of DEM, and is very efficient in the contact resolution especially for rounded particles as demonstrated by the validation cases. Nevertheless, for particles with sharp edges (e.g. a cube), MS-DEM loses its advantage as many small spheres are needed to mimic the edges, while significantly less elements (e.g. triangles) are required for a good shape approximation in polyhedral DEMs. Therefore the next step is to add the support of polyhedral DEM within the present software framework, so that our hybrid DEM framework has great versatility in terms of particle shape representation. For instance, the multi-sphere model is efficient for rounded particles, e.g. only few tens of spheres are necessary for a good shape approximation for ellipsoid, sphero-cylinder in the validation cases; while the polyhedral model is the best choice for particles with sharp edges, e.g. a cube can be precisely represented by only 12 triangles.

Even better, we could make the MS-DEM and polyhedral DEM to work together in a awkward situation, e.g. the mixture of spheres and cubes, because either MS-DEM or polyhedral DEM is not efficient in this case. For this purpose, we will need to develop an efficient contact algorithm for the interaction between particle's multi-sphere and surface models. An algorithm for the sphere-triangle contact problem may be one of the options. In this sense, our DEM framework could become truly hybrid in choosing appropriate particle models and contact detection algorithms for an efficient simulation of multi-phase flow.